\documentclass[aps,prl,twocolumn,groupedaddress]{revtex4}

\begin{document}


\title{\bf A summary on two new algorithms\\
for Grover's unsorted database search problem}


\author{Bin Shang}
\email[]{binshang@bit.edu.cn}
\affiliation{School of Computer Science \& Technology,Beijing
Institute of Technology, No. 5, Zhongguancun Nandajie, Haidian
District, Beijing 100081, P. R. China}


\date{\today}
\begin{abstract}
In this summary we discuss two new algorithms for Grover's unsorted
database search problem that claimed to have reached exponential
speedup over Grover's original algorithm. One is in the quantum
setting with ``power queries" that allow for exponential reduction
in the number of queries over Grover's original algorithm with ``bit
queries". The other is to use ``dubit queries" on a duality computer
- a new computing model uses a quantum system's wave-particle
duality, which is able to achieve even greater computing power and
better capability than existent quantum computers we have been
discussing. We discuss the shortages and difficulties of both
schemes as well.
\end{abstract}

\maketitle

\section{Introduction}
Grover's unsorted database search problem has been extensively
studied in the past decade for its wide use and great significance
in computational theory. Grover's problem is related to many famous
and important problems including the satisfiability problem, the
traveling salesman problem,the minimization problem of choosing the
smallest number out of N real numbers, the Boolean mean problem, the
integration problem, and the Sturm-Liouville eigenvalue problem. If
any one of those problems can be effectively solved it will be a
great progress to quantum computation and computer science.\\
Since 1996 people have worked on Grover's algorithm from different
aspects.~Some authors have tried to solve the search problem with a
polynomial algorithm. But people never achieved, without consumption
of exponentially more computing resources and spaces, an exponential
speedup over classical algorithms or Grover's original algorithm, as
Shor's factoring and discrete logarithm algorithms did. Nearly all
those who claimed to have improved Grover's algorithm to an
exponential speedup have been wrong.\\
\section{The Algorithm on a Quantum Computer with Power Queries}
Unlike such methods that people used before, Papageorgiou et al,
during these two or three years, have found a new way for the search
problem, which is possible to be implemented. They use ``power
queries" (``power query'' is defined in\cite{1}), which have been
applied in phase estimation algorithms and Shor's factoring
algorithm, instead of ``bit queries" applied in Grover's original
algorithm. They first\cite{1} studied the complexity of the
Sturm-Liouville eigenvalue problem on a quantum computer using power
queries, and showed that their algorithm had achieved an exponential
reduction in the number of power queries over that of bit queries .
Then\cite{2} they showed that some famous NP-hard and NP-complete
problems including Grover's Search Problem can be reduced to the
Sturm-Liouville eigenvalue problem in the quantum setting with power
queries . They solved Grover's search problem---to find single
marked state from an unsorted database with $N=2^n$ states in
all---with probability $1-\delta$, using $n \log\delta^{-1}$ power queries and $n$ qubits.\\
However, contrary to Shor¡¯s algorithm, it is doubtful that whether
power queries used in Papageorgiou et al's search algorithm can be
implemented by a number of existing elementary quantum gates that is
polylog in $n$. A recent paper\cite{3} discussed the relation
between Grover's and Shor's algorithms, and showed that Grover¡¯s
algorithm can be viewed as a quantum algorithm which solves a
non-abelian hidden subgroup problem (HSP) but the standard
non-abelian quantum hidden subgroup algorithm can not find a
solution to this particular HSP. That work would be helpful to
studying on how to build a new method to solve Grover's search
problem with power queries without the exponentially increasing
difficulties in constructing the quantum gates. If such method is
found, we can affirm that quantum computers are able to solve
efficiently NP problems.\\
\section{The Algorithm For Grover's Search Problem On a Duality Computer}
\subsection{1.Introduction}
Almost simultaneously, Long has spent more than three years thinking
on his brand new computing model using the wave-particle duality of
quantum systems: the duality computer\cite{4}\cite{5}. From a
physician's view, he considered the principles of quantum
interference between both quantum particles and quantum systems. He
stood on convictive physical basis, by generalizing series of
principles of quantum interference, and proposed two implementations
for the duality computer.\\
A duality computer is such a computing machine that uses both
particle and wave property of quanta.Analogous to qubit,Long uses
``dubit" (duality bit) as the basic information unit. Such is a
sequent process of a computation on a duality computer (duality
computation for short): first, we make the initial quantum states of
the quantum system for the computation;second, we divide the quantum
wave of the system into multiple paths (i.e. sub-waves) using a
setting called quantum wave divider (QWD for short); third, we
perform different quantum gate operations on corresponding paths;
fourth, we recombine all the sub-waves to get the resultant wave
using another setting called quantum wave combiner (QWC for short);
at last,we measure the readouts of the computation.\\
\subsection{2.Complexity}
Now let us see the complexity of Long's algorithm for Grover's
unsorted database search problem---to find single marked state from
an unsorted database with $N=2^n$ states in
all---with a single query on a duality computer.\\
Here we ignore all errors that may occur. It is shown in \cite{5} that we can find the marked item with certainty.\\
(1) Dubit complexity\\
The algorithm can be implemented using only dubits.So the dubit
complexity is  $\log N=n$.\\
(2) Query complexity\\
Only one query is needed for one dubit for the problem. In
randomized cases, the lower and upper bounds of the algorithm are
$\Omega(\log N)$ and $\mathcal {O}(\log N)$. So the query complexity
is $\Theta(n)$.\\
Obviously, Long's algorithm has achieved an even greater speedup
over Papageorgiou et al's.
\subsection{3.Difficulties}
Although the proposal of duality computer is exciting, and
algorithms on a duality computer can achieve outstanding efficiency
on the unsorted database search problem, it has both theoretical and
practical difficulties.\\
Unlike in a quantum computer, Long used the generalized interference
principles which are not currently accepted by most physicians. If
the hypothesis that interference will occur to quantum systems with
several constituent particles, no matter loose or bound as Long
proposed, is proved wrong, the duality computer can be of only
theoretical interest. \\
Besides, it is unknown yet whether we can in the near future succeed
in constructing such a computer that is able to subtly operate
dubits' (such as photons) both particle and wave properties. The
difficulty may be much more than what we are suffering in
constructing (optical) quantum computers.
\section{Conclusion}
We showed two completely different algorithms for Grover's unsorted
database search problem in quantum settings with power queries and
in duality computers. We viewed the complexities of both algorithms,
and found either algorithm achieves an exponential speedup than any
former algorithms and either way may be a great innovation on
quantum computation.\\
We studied shortcoming of the two schemes as well. Both methods are
in theory, and not yet implemented in practice, or perhaps will
never be. However, in the author's opinion, studies on both methods
are meaningful. Even if they are of only theoretical value, we may
find other helpful ways related to them to solve Grover's search and
other NP hard problems. And if only one of them can be implemented
some day, we will have progress a great step in computing.

\end{document}